# Rapid Spreadsheet Reshaping with Excelsior: multiple drastic changes to content and layout are easy when you represent enough structure

*Jocelyn Paine, Emre Tek, Duncan Williamson*
[www.j-paine.org/](www.j-paine.org/) *and* [www.spreadsheet-factory.com/](www.spreadsheet-factory.com/)
[popx@j-paine.org](popx@j-paine.org)

**ABSTRACT**

*Spreadsheets often need changing in ways made tedious and risky by Excel. For example: simultaneously altering many tables' size, orientation, and position; inserting cross-tabulations; moving data between sheets; splitting and merging sheets. A safer, faster restructuring tool is, we claim, Excelsior. The result of a research project into reducing spreadsheet risk, Excelsior is the first ever tool for modularising spreadsheets; i.e. for building them from components which can be independently created, tested, debugged, and updated. It represents spreadsheets in a way that makes these components explicit, separates them from layout, and allows both components and layout to be changed without breaking dependent formulae. Here, we report experiments to test that this does indeed make such changes easier. In one, we automatically generated a cross-tabulation and added it to a spreadsheet. In the other, we generated new versions of a 10,000-cell housing-finance spreadsheet containing many interconnected 20×40 tables. We varied table sizes from 5×10 to 200×2,000; moved tables between sheets; and flipped table orientations. Each change generated a spreadsheet with different structure but identical outputs; each change took just a few minutes.*

## 1. INTRODUCTION

You are a spreadsheet consultant. Your client has given you a ten-thousand-cell spreadsheet: a model, let it be, of how the British economy will grow over the next 40 years. The client tells you that 9,000 cells lie within tables. Some tables are two-dimensional: they segment the economy into 20 market sectors, and record, for each sector, such year-by-year characteristics as employment, inflation, and growth. Each table holds $20 \times 40 = 800$ cells. Other tables are one-dimensional: they record year-by-year values aggregated over all sectors, or sector-by-sector values aggregated over all years. Of the remaining 1,000 cells, 500 contain table headings. And of the 500 cells that aren't text or table, 400 belong to blank columns and rows between tables; 100 are user-modifiable assumptions.

Your task is to make the tables resizable, so that the client can produce, on demand, a super-large spreadsheet capable of 100 years by 75 sectors; or, to save memory and avoid scrolling, a mini-version a mere 10 years by 5 sectors. Or any other size they fancy. By the way, it would be nice if we could rearrange any or all of the tables so that years run across and sectors down, instead of years down and sectors across. We'd quite like to be able to move tables from sheet to sheet. Oh, and we sometimes need bespoke models, so can you make it possible for us to replace any table with a new one, custom-built.

I hope readers will realise this is not a contrived problem, but is typical of what developers would like to do with their spreadsheets, had they the tools. Indeed, it is isomorphic to a real-world commercial spreadsheet to which two of us — Jocelyn Paine and Emre Tek — applied Excelsior, as I shall explain.

The crux is that developers often need to change spreadsheets for very pragmatic reasons: making them easier to read; saving memory; making room for extra columns or rows; adding new reports; adapting to the needs of special clients. And the developer can often think about a change very concisely: "Dear Excel, please resize each sector-by-year table to have 30 sectors and 100 years". But because Excel knows nothing about higher-level structure such as how the spreadsheet is divided into tables, the developer has no way of saying this, and must perform the change cell by tedious cell. — The book The Making of "The Hitchhiker's Guide to the Galaxy" [Stamp, 2005] explains how the set designers gave the Vogons a language in which every number had to be written as so many 1's. This is why Vogon civil-service forms took so long to fill in, and hence the civil servants were permanently bad-tempered. Excel represents spreadsheets as if it were a Vogon.

Unlike Excel, Excelsior does know about higher-level structure — in particular, how spreadsheets are divided into components. Excelsior is the result of a long research project into reducing spreadsheet risk. Part of its motivation is the hypothesis that creating and changing spreadsheets will be safer and easier if they are represented in a way that makes their high-level structure explicit. More precisely: in a way that describes how they are divided into components, isolates this from layout, and allows both components and layout to be changed without breaking dependent formulae.

**1.1 Content of This Paper**

In this paper, we demonstrate that this is so. Our main experiment, the first time Excelsior has been tested on a realistic spreadsheet, was to restructure a 10,000-cell housing-finance spreadsheet used by social-housing consultants Weedon Grant. This spreadsheet contains about 60 interconnected 20×40 tables. We generated new versions of this spreadsheet, with table sizes varying from 5×10 to 200×2,000. We also moved tables between sheets; and flipped their orientations. Each change generated a spreadsheet with different structure but identical outputs; and each change took just a few minutes.

That experiment is described in Section 3. Section 4 describes a smaller experiment, in using Excelsior to programmatically generate a table and insert it into a spreadsheet to cross-tabulate its data. That is useful in its own right, and also demonstrated the advantages — this time for the advanced programmer rather than the spreadsheet developer — of working with data structures that represent spreadsheets in a high-level layout-independent fashion. Sections 2 explains Excelsior, Section 5 is our conclusion; Section 6 is the bibliography.

**1.2 Previous Work**

As already mentioned, my research has been driven by the idea that high-level structure must be represented. I have explored languages wherein spreadsheeters can describe structure and then compile their programs into Excel; but, because most will prefer developing in Excel, I have also explored how to recover structure from existing Excel spreadsheets. I call this "structure discovery". This work is described in [Paine, 2001; 2004(a); 2004(b)].

Rapid Spreadsheet Reshaping with Excelsior: Paine, Tek & Williamson

A final paper, [Paine, 2005], describes Excelsior, in which I settled upon a notation and a relatively efficient implementation for use on large, realistic, jobs. The big achievement here — of which I am particularly proud — is to have produced the world's first ever program for modularising spreadsheets, endowing spreadsheeting with the same benefits that modularisation already confers upon other branches of computer science and engineering.

**2. BACKGROUND: INTRODUCTION TO EXCELSIOR**

This section explains enough of Excelsior to make sense of the rest of the paper. A fuller account is given in [Paine, 2005].

**2.1 Spreadsheets Are Sets Of Equations**

Excelsior regards a spreadsheet as a system of equations relating the elements of tables. Thus, suppose the one-dimensional tables `Builds` and `Demolitions` give the number of dwellings built and demolished per year, and `NewStock` gives the net new dwellings per year. Then we could write two year's calculations as what is clearly a set of two equations:

```
  NewStock[ 2000 ] = Builds[ 2000 ] - Demolitions[ 2000 ]
  NewStock[ 2001 ] = Builds[ 2001 ] - Demolitions[ 2001 ]
```

Tables can have more than one dimension. Should we want more than one dwelling type, we could make it a second dimension to our tables:

```
  NewStock[ 2000, 1 ] = Builds[ 2000, 1 ] - Demolitions[ 2000, 1 ]
```

To do the job of repeated formulae in Excel, Excelsior has an "all" construction which generates equations relating all elements of tables. Thus, Excelsior will expand the following into as many equations as there are elements of `Builds`, `Demolitions` and `NewStock`:

```
  NewStock[ all y, all dt ] = Builds[ y, dt ] - Demolitions[ y, dt ]
```

I call the above a "quantified equation", by analogy with the universal quantifier in logic, which ranges over all values of a variable. Excelsior also permits bounds to be specified for quantified variables. This is particularly useful for separating the calculation of initial values, for example for the first time point in a range, from that of subsequent values:

```
  NewStock[ 2000,    all dt ] = 0
  NewStock[ y>2000, all dt ] = Builds[ y, dt ] - Demolitions[ y,dt ]
```

**2.2 Object = Tables + Equations**

An Excelsior "object" is a collection of table declarations together with equations. (I took the word "object" from the mathematics on which Excelsior was based; it has no direct connection with object-oriented programming, though ultimately both are related.) Here is an object:

```
{#
  NewStock[ 2000:2010, 1:20 ],
  Builds[ 2000:2010, 1:20 ],
  Demolitions[ 2000:2010, 1:20 ]
|
  NewStock[ all y ] = Builds[ y ] - Demolitions[ y ]
#}
```

In keeping with Excelsior's sets-of-equations semantics, I adapted this notation from the mathematical notation for sets. Mathematics encloses sets in the brackets { and }; Excelsior encloses sets of equations in the brackets {# and #}. (Choice of programming-language notation is always a losing battle between the elegance of handwritten or printed mathematics and the eternal disdain of computer keyboard designers for anything other than ASCII; and there are never enough brackets. However, the hash # looks like part of a spreadsheet grid, so the symbols {# and #} seemed a reasonable compromise to bracket equations that will eventually become a spreadsheet.) The vertical bar | is set theory's notation for "such that". With this key, the above code can be read as:

```
The tables NewStock, Builds and Demolitions,
each of which runs from 2000 to 2010 and 1 to 20,
such that
for all y, NewStock[ y ] = Builds[ y ] - Demolitions[ y ].
```

## 2.3 The "Union" Operator Glues Objects Together

The operator ∪ combines objects by forming the union of their tables and sets of equations:

```
{# a[1:1], b[1:1] | a[1]=b[1] #} ∪
{# a[1:2], b[2:3], c[] | c[]=a[2], a[2]=a[1] #} =
{# a[1:2], b[1:3], c[] | a[1]=b[1], c[]=a[2], a[2]=a[1] #}
```

I shall not need union in Section 2, but mention it now because I shall use it in later sections.

## 2.4 Resizable Tables Are Implemented Via Object-Valued Functions

To make tables of adjustable size, we define functions that take size parameters and return objects. The bounds of tables in these objects can depend on the size parameters. Thus the example above could be generalised to:

```
let model( StartYear, EndYear, NumberOfDwellingTypes ) be
  {#
    NewStock[ StartYear:EndYear, 1:NumberOfDwellingTypes ],
    Builds[ StartYear:EndYear, 1:NumberOfDwellingTypes ],
    Demolitions[ StartYear:EndYear, 1:NumberOfDwellingTypes ]
  |
    NewStock[ all y ] = Builds[ y ] - Demolitions[ y ]
  #}
```

The example's first line introduces a function definition, including the names of the function's arguments.

Because `model` is a function, we can call it. We use the same notation for function calls as almost every other programming language:

```
model( 2000, 2040, 20 )
```

And by changing the numbers, we can generate objects having tables of any size we want.

**2.5 Mapping An Object's Tables Onto Worksheets**

One thing remains before we can make these objects into spreadsheets: telling Excelsior how tables are to be represented as spreadsheet cells. Should years run horizontally or vertically? At which cell should each table start; on which sheet; and with what row and column captions? I handle this by making Excelsior regard a spreadsheet also as a set of equations, equivalent to an object in which each table is a separate worksheet. Converting a general object to a spreadsheet is then a mapping of coordinate systems in which each table is mapped onto one or more of the worksheets.

To transform coordinates, we use Excelsior's `mapping` function. This takes three arguments: an object, a source range, and a target range. It returns a new object that is the same as its first argument, except that its equations have been rewritten so that all references to cells in the source range are replaced by references to corresponding cells in the target range.

As an example, suppose `Obj` to be an object that contains a table named `Lettings`. Here is code that maps this table onto cell D8 of sheet `Lets`:

```
Obj mapping Lettings to Lets!D8 by yx
```

The `by yx` specifier means that the table's first dimension is mapped onto the sheet so that it runs downwards ("y"), while the second dimension runs across ("x").

**2.6 Structure-Discovery Maps Worksheets To Tables, Making Them Understandable**

That worksheets are two-dimensional tables makes it easy to map from spreadsheets as well as to them. Doing so is essential in order that we can convert existing spreadsheets into Excelsior objects wherein groups of related cells are explicitly named as tables. If cells are grouped correctly, and the tables given sensible names, then the object will be much easier to read than the spreadsheet. This has been a major theme in my research. Its importance is evident by noticing how intelligible are the examples at the start of Section 2.

Moreover, we can then use the Excelsior object created from the spreadsheet to generate different but structurally similar spreadsheets.

**3. RAPID RESTRUCTURING OF A LARGE FINANCIAL SPREADSHEET**

**3.1 Introduction**

Generating different but structurally similar spreadsheets brings us to our (Emre Tek and Jocelyn Paine's) experiment in doing exactly that with Weedon Grant's housing-finance spreadsheet. This was, as it happened, the first stage — the feasibility test — of a commercial project, in which we would use Excelsior to make an assortment of financial spreadsheets easier and faster to restructure and customise than with Excel. But it is also an interesting, and we believe, valuable piece of spreadsheet research, worth reporting as such.

This work had two stages. Firstly, we converted the original spreadsheet into Excelsior notation, by structure discovery. We then had to modify this so it could generate variants of the spreadsheet. This is described in Section 3.2.1. We also had to devise some Excelsior notation for describing a spreadsheet's layout, in a way independent of the size of its components and also easy for novice users to edit. This is described in Sections 3.2.2 and 3.2.3. Our notation is new since [Paine, 2005], though it is based on the operators described there.

In the rest of this introduction, we explain the needs of the housing-finance developers we hope to help, and then present our research hypothesis. Section 3.2 then describes the implementation; and Section 3.3, our evaluation. Section 3.4 is a note on how Excelsior's modularisation facilities, which are not the main topic of this paper, can also help.

**Why The Stock-Model Spreadsheet Needs Such Drastic Editing**

From here on, we shall refer to the housing-finance spreadsheet as the "stock-model" spreadsheet, because it models housing stocks. We first need to explain this.

Weedon Grant is a management consultancy working in social housing and regeneration, advising councils and housing associations. They refine their spreadsheet-based business planning models to match the needs of their clients. The backbone of their business planning is the housing stock model. In structure, this is like the 10,000-cell economic model with which I opened, except that instead of economic growth, it forecasts year-by-year stocks of, and both income and expenditure derived from, a housing association's houses and flats. The stock model's tables are indeed 20 wide, 40 deep, or both, and do encompass roughly 9,000 cells. Instead of market sector, the length-20 dimension represents different types of dwelling, for example flats, houses or bungalows with one, two or three bedrooms.

Different housing associations have different numbers of dwelling types, and need to forecast over different numbers of years. For this and the other reasons suggested in the introduction, it would be extremely useful if we could resize all the tables on demand. Naturally, the tables' captions should move with the tables; and the general layout, including single-cell inputs and outputs near tables, and the rows and columns between tables, should be preserved.

We may also want to change layout more radically; for example, some clients might prefer years to run across rather than down, or might want certain non-essential tables to be collected onto a single sheet and hidden from view. If we could do these things, it would be a bonus.

**Research hypothesis**

Stated more formally, we wanted to test the hypothesis that: given a large and complicated (and commercially used) spreadsheet, we could automate the task of making numerous drastic but related changes such as the table resizing described above, in a way easier, faster, and safer than doing so in Excel.

Please note that we did not try to reproduce properties such as cell formats and colours, input menus, and charts. We are taking things one step at a time: our only concern was to reproduce the correct formulae. Having managed that, we are now making Excelsior handle these other properties.

### 3.2 Implementation

**Structure-discovering and rebuilding the stock model**

This section describes how we converted the original spreadsheet into Excelsior notation (an object definition) by structure discovery, and then modified the result (to an object-valued function) so it could generate variants of the spreadsheet. We followed these steps:

1. In each of the stock-model spreadsheet's tables, there are runs of formulae, identical except for the cells they operate on. To find these, I coded a repeated-formula detector. This works similarly to that described in the *Loose ends* section of [Paine, 2004(b)], but detects runs over an arbitrary number of dimensions, not just one.

    Such runs may span an entire table; or perhaps part of a table, should the first row or column have been created differently from others. So they, plus visual inspection and (sketchy) information about the stock model's workings, gave me those ranges likely to be tables. I guessed suitable names for these tables from captions nearby; then coded the result as arguments to Excelsior's `mapping` function.

    Call this `M`. Then we can say that `M` is a set of tuples ( `w`, $c_l$, $c_u$, `t`, $b_l$, $b_u$ ). In each tuple, the range `w!`$c_l$`:`$c_u$ is to be mapped to the Excelsior table `t[`$b_l$`:`$b_u$`]`, where the name `t` has been guessed from an appropriate caption near this range.

2. I then applied this mapping `M` to the stock-model spreadsheet, generating an Excelsior object `E`.

3. Although `E` contains named tables rather than worksheets, it has the same number of cells as the stock-model spreadsheet, and corresponding groups of repeated formulae. We could list all these with Excelsior's `show` function. However, the listing is not useful, because it contains about 10,000 lines of tediously repetitious equations.

4. Therefore, I applied the repeated-formula detector again, but to `E` rather than the stock-model spreadsheet. Call the result $E_{compressed}$. In $E_{compressed}$, runs of repeated formulae had been compressed into single quantified equations. This gave us a much smaller listing, 450 equations in size.

5. Some of these equations calculated strings (text) rather than numbers. In a few of these, other calculations depended on the resulting text; in the rest, the text was not used in further calculation, and I assumed it to be captions. I moved these to a separate file. Symbolically, we can represent this by saying that we split the object $E_{compressed}$ into an object `A` containing only annotations and an object `C` containing only calculations.

6. I then wrote a layout spreadsheet `L` of the kind described in the Section on "Using Spreadsheets for Visual Specification of the Layout of Resizable Spreadsheets". This contained the annotations from `A`, and the table positions from the mapping `M`.

7. We then edited $E_{compressed}$ and `L` to change some of the table names. Because guessed from captions, they weren't all as informative as we wanted.

8. I then tried regenerating the original stock-model spreadsheet from the calculations-only specification `C` and the layout spreadsheet `L`. At this stage, I discovered that I had wrongly guessed the positions of some tables and whether some of the text cells were really only captions. We therefore had to repeat some of the previous stages until we got the tables correct.

9. After this, we were able to generate a spreadsheet identical in layout and calculations to the original stock-model spreadsheet. When given the same inputs, it calculated the same outputs.

10. Next, we edited the calculations-only specification `C` again, replacing constant table sizes by named variables. We added an Excelsior function-definition line, using the variables as its formal parameters. This gave us an object-valued function `F`.

11. We tested `F` by calling it with the stock-model spreadsheet's table sizes as arguments. This again regenerated a spreadsheet identical in layout and calculations to the original.

12. *However*, when we called `F` with other values of the size parameters, we were able to generate stock-model spreadsheets in which all the tables had been resized.

    We tested these by supplying them with the same inputs as the original, and again they gave the same outputs. We also inspected intermediate calculations to check that their results were as in the original. (In doing this, we took care not to make the tables so small that there was not room for the needed inputs, outputs, and intermediates.)

    We tried this with sizes up to a monster 200 dwelling types and 2,000 years. It is unlikely that any piece of social housing built today would remain standing so long; but it did prove that sheer size was not a problem.

    We noted however that, although it took only a few minutes to make the changes, the time taken for Excelsior to output the XML file containing the generated spreadsheet was worse than linear, up to about 20 minutes for the largest tables. This was because Excel rejects the XML unless cells are sorted by row and then column within each worksheet, and so Excelsior has to sort them. Profiling showed the sort to be taking most of the time. We are looking for ways to circumvent this.

    (In requiring cells to be sorted, Excel is being deliberately obtuse. The XML elements for cell within row and for row within sheet contain attributes for x-position

and y-position respectively, and so semantically, the XML would be correct whatever order cells are output in, as long as two cells never have the same position.)

13. Furthermore, by editing the layout spreadsheet `L`, we generated wildly different versions of the stock model spreadsheet, with rows and columns inserted, rows and columns deleted, tables flipped, tables moved within and across sheets, sheets merged, sheets split. Each change took as long as was needed to edit the layout spreadsheet in Excel and then invoke the Excelsior compiler: generally a minute or two.

14. At this stage, we uncorked a bottle of champagne. Well, we would have, but work called, and I needed to get back from Emre's flat in London to Oxford. Perhaps after the EuSpRIG 2006 demo.

**Using Formats For Textual Specification Of The Layout Of Resizable Spreadsheets**

Section 2.5 explained the `mapping` function for mapping tables to worksheets. The trouble with it is that if tables can vary in size, their origins will probably move. And so will the position of captions. For example, suppose the stock model must have a `Sales` table to the right of `Lettings`, separated from it by a blank column. Then `Sales`'s origin will obviously depend on the number of columns in `Lettings`. So will the position of any text used as column headings.

I did take this into account when designing Excelsior, by allowing the cell addresses in a mapping to be calculated from size variables. A cell address in Excelsior is a structure containing a sheet name and a two-dimensional vector. Expressions like `Lets!D8` denote constant cell addresses. But general expressions are allowed anywhere that a constant is, so we can calculate cell addresses using vector arithmetic:

```
Lettings to Lets!D8 by yx
Sales to (Lets!D8) + vector(NumberOfDwellingTypes+1,0) by yx
```

Although this notation is general enough to describe any layout at all, it turned out to be inconvenient for the stock model. This has a lot of adjacent tables, and almost everything on a worksheet would need to change position if these were resized. This would entail coding a lot of cell-address calculations, with consequent risk of errors such as typos and "off-by-one" mistakes. Since Excelsior is intended to reduce risks, not increase them, this isn't acceptable. One solution I considered was to allow table-relative specifications such as:

```
Lettings to Lets!D8 by yx
Sales to top_right(Lettings) + vector(1,0) by yx
```

Trying this out on the stock model, problems became apparent. Firstly, this still needed too much typing. Secondly, if the position of one table depends on that of another, and that upon a third, … , we end up with a huge web of dependencies. This web is fragile: one table depends on many others, and moving or deleting any one of these may snap many strands in the web.

After experimenting with various possibilities, I found one that bypassed these problems. The idea is to borrow from the notion of "format". As used in most programming languages, a format is a template or picture of text to be displayed. The format specifies parts of the text that won't change, but also has holes into which variable text can be slotted. For example, in SWI Prolog, the following command inserts values for Fahrenheit and Celsius into a string:

```
format( 'The Fahrenheit equivalent of ~w°C is ~w°F'
      , [ DegreesC, DegreesF ]
      ).
```

Applying the format idea to Excelsior, I invented a notation that looked like this:

```
row( [ Lettings by yx, skip, Sales by yx ] ) @ Lets!D8
```

This describes one worksheet's layout. The worksheet contains three things in a row starting from `Lets!D8`. These are: the table `Lettings`; a blank column, denoted by `skip`; and the table `Sales`. Both tables are to be laid as indicated by the `yx` specifier.

Note that when I say "row" here, I mean a row of tables, not a row of cells. If either table is more than one cell deep, Excelsior will insert blank cells under each item until it is the depth of the deepest. The effect is to give a row of items all of which are the same depth.

To implement this, imagine working along the list keeping track of where each item will start on the worksheet, in terms of the widths of the items before it. The start of the first item is given by the `@Lets!D8`. This item's width can be calculated from its dimensions when laid out as `yx`. This gives the position of the blank column denoted by `skip`; and so on.

It is for ease of explanation that the above example uses a row of items. In general though, items need to be arranged vertically as well as horizontally. This is done with the `grid` format, thus:

```
grid( [ [ 'STOCK MODEL']
      , [ skip(0,3)      ]
      , [ 'Years'       , skip, 'Lettings'    , skip, 'Sales'    ]
      , [ Years by y    , skip, Lettings by yx, skip, Sales by y ]
      ]
    ) @ Lets!A1
```

The argument to `grid` is a list of lists. Inner lists represent rows of items. Elements in corresponding positions of the outer list represent columns of items. Items in quotes represent text, and will be given one cell each.

The symbol `skip` now gets two arguments, the first being a number of columns to skip, the second being a number of rows. Thus `skip(X,Y)` denotes a box of empty cells *X* long and *Y* deep. On its own, `skip` is short for `skip(1,0)`.

To generate a spreadsheet from such a format, the tops of items in the same row are horizontally aligned; the items are padded beneath with blank cells to the depth of the deepest. The left-hand sides of items in the same column are vertically aligned; the items are padded on their right with blank cells to the width of the widest. In this example, this would put the text `'STOCK MODEL'` in A1; nothing in rows 2, 3 and 4; text in row 5; and so on. Adjacent tables will be separated by blank columns, and each table placed below its encaptioning text.

**Using Spreadsheets For Visual Specification Of The Layout Of Resizable Spreadsheets**

Grid formats were adequate to lay out an Excelsior version of the stock model so as to replicate the original spreadsheet. But although more convenient than the `mapping` function, they were still not convenient enough. Moving a table often required a lot of text editing, and it was hard to see when elements in a grid format were vertically or horizontally aligned.

To overcome this, I hit upon the idea of depicting grid formats as a spreadsheet. Thus, the format

```
grid( [ [ 'STOCK MODEL']
      , [ skip(0,3)     ]
      , [ 'Years'       , skip, 'Lettings'   , skip, 'Sales'   ]
      , [ Years by y    , skip, Lettings by yx, skip, Sales by y ]
      ]
    ) @ Lets!A1
```

would be equivalent to the spreadsheet

| 'STOCK MODEL' |      |             |      |          |
|---------------|------|-------------|------|----------|
| skip(0,3)     |      |             |      |          |
| 'Years'       | skip | 'Lettings'  | skip | 'Sales'  |
| Years y       | skip | Lettings yx | skip | Sales y  |

My research over the past few years has aimed to replace or supplement spreadsheets by textual specifications, so it must seem perverse that I am now doing the reverse. However, the layout spreadsheets have a very different use from that of normal spreadsheets. They don't calculate results, but merely depict the relative positions of certain items. There are no cell references to be mistyped. And Excelsior will detect errors such as duplicate table names and tables whose names occur in the layout spreadsheet but not the stock-model specification or vice-versa.

**3.3 Evaluation**

To recap, we wanted to test the hypothesis that: given a large and complicated (and commercially used) spreadsheet, we could automate making numerous drastic changes in a way faster, easier and safer than doing so in Excel.

We are very very pleased. Once structure discovery was completed and the layout spreadsheet written, we made vast changes to size and layout just by changing the size parameters or the layout spreadsheet and rerunning Excelsior. It was much faster and easier than changing the original stock-model spreadsheet would have been. We didn't try to measure error rate. However, it does seem that changing one parameter must be safer than, e.g. making big alterations to 60 tables and all their dependent formulae.

Every new tool does, of course, provide new opportunities for error. In this case, it was the structure discovery, which was only semi-automatic and which required care in editing the Excelsior code and checking against the original spreadsheet. It also took quite a time: about 2 days in total (although that was interspersed with mods to Excelsior).

For the experiment described here, that did not worry us. We used structure discovery because we needed to replicate the original stock model exactly. However, in future work on Weedon Grant's spreadsheets, we shall recode from scratch in Excelsior. In other research, we shall automate structure discovery further.

Properties such as cell styles were lost by the structure discovery stage. But as explained in Section 3.1, we were not trying to reproduce these. We are now looking into this.

I want to end with a slogan which forcibly suggested itself to me when we changed the layout spreadsheet and watched Excelsior generate new stock-model variants, and yet saw all their results stay exactly the same because of the compensating changes Excelsior made to all formulae that depended on this layout. For many spreadsheet jobs:

> **THE MEANING OF A SPREADSHEET SHOULD BE INVARIANT OVER CHANGES TO LAYOUT.**

That is what we have provided.

### 3.4 A note on modularization

I finish by noting that we used Excelsior not only because it could rapidly restructure spreadsheets, but also because we need modularity. Some stock-model users will want bespoke versions, containing the same "core" as all the other versions, but different additional features.

Symbolically, we can write this by saying that one client wants `Core ∪ FeatureA`; another wants `Core ∪ FeatureB`; and so on.

Actually, things will be more complicated. We know that `Core` will contain the resizable tables that are the main topic of this section. So it will actually be an object-valued function that gets called with appropriate size arguments, e.g.

```
Core( 2000, 2040, 20 )
```

But the additional features will also contain such tables; and we must obviously keep their sizes in step with the tables in the core. We shall do so by implementing the additional features as object-valued functions too. Then we shall be able to generate one client's model as, say,

```
Core( 2000, 2040, 20 ) ∪ FeatureA( 2000, 2040, 20 )
```

while the other client's might be:

```
Core( 2005, 2030, 40 ) ∪ FeatureB( 2005, 2030, 40 )
```

We shall refine these ideas as the project continues.

### 4. GENERATING CROSS-TABULATIONS

### 4.1 Introduction

I now move to the second experiment, to programatically generate a cross-tabulation and insert it into a spreadsheet. In the rest of this introduction, I explain this spreadsheet, why we want automatically-generated cross-tabulations, and why pivot tables aren't good enough. I also present the research hypothesis. Section 4.2 is background on cross-tabulations; Section 4.3 describes the implementation; Section 4.4 evaluates it.

**About The Spreadsheet To Be Tabulated**

This experiment was suggested by Duncan Williamson, inspired by an accounting spreadsheet he wrote for a company which makes ceramic tiles by kiln-treating clay. The spreadsheet generates reports showing manufacturing cost breakdowns; these reports are cross-tabulations, calculated from inputs giving quantities and unit costs of such things as: clay, barium carbonate, and other raw materials; water and gas; pallets and other equipment; wages and pensions of skilled and unskilled staff. Here is an example tabulation:

|  | Direct or Indirect | Stage | Cost element | Total |
|---|---|---|---|---|
|  | Direct | Kiln drying | Barium Carbonate | 1,999.01 |
|  |  |  | Unprocessed clay | 111,945.86 |
|  |  | Total |  | 113,944.07 |
|  | Indirect | Kiln drying | Electricity | 2,234,567.89 |
|  |  | Total |  | 2,234,567.89 |

**Why Not Use Pivot Tables?**

Duncan's spreadsheet does its tabulations with pivot tables. But these have disadvantages, which we wanted to overcome. The problem is that Excel makes the user specify pivot-table structure — what to tabulate against what — via a graphical interface. For each pivot table, Excel displays a menu of possible items to tabulate; the user must select from these until the structure of the table is fully determined. When there are many similar tables, this becomes a tediously repetitive preliminary to obtaining one's results. Could I instead program Excelsior to take the spreadsheet's inputs together with a declarative specification of table structure, generate all the tables from these, and insert them into the spreadsheet?

**Research Hypotheses**

Stated more formally, I wanted to test the hypothesis that: it is possible to automate the task of generating many similar tables from a declarative specification of their structure, then inserting them into an existing spreadsheet to cross-tabulate its data.

Because of lack of time (much of Excelsior has been a spare-time unfunded project), I had to cut this down a bit. I looked into notations for specifying tables, but shan't describe them here. There also wasn't time to experiment with generating all the tables needed to report on the original spreadsheet: this is quite complicated, and there are subtle variations between them. So I tried automating the task of generating just one table from a representation of its structure hard-wired into my program, then inserting it into a spreadsheet to cross-tabulate its data.

As with Section 3.1, I was not trying to preserve the original spreadsheet's cell formats, cell colours, and so on.

A key theme of this paper is the importance of high-level layout-independent representations of a spreadsheet. As explained in Section 1.1, I believe that a library of such representations and operations on them would be useful to programmers needing to write programs to manipulate spreadsheets. So a second objective of this experiment was to test the hypothesis that: not merely is it possible to automate the task of generating and inserting tabulations, but that it is easy to do so using Excelsior's data structures.

**4.2 Background: Explanation Of Cross-Tabulation**

If you are unfamiliar with cross-tabulation, you can find a a good explanation in the section about *Cross tabulation and stub-and-banner tables* of [StatSoft]. In general, a cross-tabulation tabulates the possible values or ranges of values of variables against one another, showing in each cell the frequency with which a particular unique combination of values occurs.

I shall illustrate the concept of cross-tabulation, and our algorithm, with a simple spreadsheet written to explain pivot tables to novices. This is also the spreadsheet I used in my experiment. It was written by Harald Staff of Pearson Software Consulting, and is available on the Web as [Staff]. Here is part of it:

|    | A     | B    | C    | D      |
|----|-------|------|------|--------|
| 1  | Who   | Week | What | Amount |
| 2  | Joe   | 3    | Beer | 18     |
| 3  | Beth  | 4    | Food | 17     |
| 4  | Janet | 5    | Beer | 14     |
| 4  | Joe   | 4    | Beer | 12     |
| …  |       |      |      |        |
| 24 | Janet | 4    | Car  | 17     |
| 25 | Janet | 5    | Food | 12     |

Suppose as an example that this spreadsheet shows how much each person has spent on various goods each week. The variables are then person, week number, kind of good, and amount spent, which the spreadsheet calls `Who`, `Week`, `What`, and `Amount`. Let's take a simple case, a two-dimensional cross-tabulation of `Who` against `What`. Then the possible values of `Who` are `Beth`, `Janet` and `Joe`. The possible values of `What` are `Beer`, `Car` and `Food`.

Our cross-tabulation will have two discrete dimensions corresponding to these two sets of values. Each cell represents a pair of values such as `Beth/Beer` or `Janet/Car`, and will record how often that pair occurs: that is, how many rows it occurs in. (In this case, we would probably prefer to record the amount of expenditure each pair represents, weighting it with the appropriate value for `Amount`. I'll ignore that here.) The table will look like this, but with the blank cells filled in with appropriate counts:

|  | Beth | Janet | Joe |
|---|---|---|---|
| Beer |  |  |  |
| Car |  |  |  |
| Food |  |  |  |

**4.3 Implementation**

All the work described in this section was coded in SWI Prolog. The original spreadsheet, the tabulation to be added, and other components such as the value-combiner range of Section "Counting Pairs Of Values" were represented by the data structures that I use for Excelsior objects, operated upon by predicates implementing operations such as union.

**The Cross-Tabulation Algorithm**

To generate cross-tabulations, I devised the following algorithm:

1. Read from the user a specification of the ranges to be tabulated against one another. In this experiment, the specification was hard-wired into the program.

2. Get the values in each range, remove duplicates, and sort into alphabetical order. The resulting sets of values will be the dimensions of the cross-tabulation table.

3. Create an Excelsior object containing one table. This object, which we shall call `T`, will hold our cross-tabulation. Its *i*'th dimension corresponds to the *i*'th set of values.

4. Into each cell of `T`, put a formula to count how often that pair of values occurs in the original spreadsheet. I say more about this in the following section.

5. The object `T` is an abstract representation of our cross-tabulation, but carries no information about how the table is to be laid out when added to the original spreadsheet. Therefore, specify a layout. This was done as an Excelsior mapping, `M`, hard-wired into the program.

6. Apply the mapping `M` to `T`, creating another object $T_s$ representing the cross-tabulation as a worksheet.

7. Insert $T_s$ into the original spreadsheet `S` by forming their union $T_s \cup S$.

Applied to Harald Staff's spreadsheet, this gave the following cross-tabulation:

|   | A    | B     | C     | D   |
|---|------|-------|-------|-----|
| 1 |      | Beth  | Janet | Joe |
| 2 | Beer | 3     | 3     | 5   |
| 3 | Car  | 0     | 3     | 1   |
| 4 | Food | 2     | 2     | 5   |

**Counting Pairs Of Values**

In step 4 of the algorithm, I said that each cell of the cross-tabulation object T gets a formula that counts how often the corresponding combination of values occurs in the original spreadsheet. I implemented this using Excel's COUNTIF function: COUNTIF( R, C ) returns the number of cells which satisfy the condition C in the range R.

Unfortunately, COUNTIF only allows one range-condition combination, whereas I needed to count co-occurrences of two values. To get round this, I generated an auxiliary "value combiner" range with as many cells as the original table had rows. Each cell in this range contained a pair of values formed by concatenating the values being tabulated from the corresponding row of the original table.

To make this clear, here are some of the formulae in this range:

|    | A |
|----|---|
| 1  | = Sheet1!$A$2 & "_" & Sheet1!$C$2 & "_" |
| 2  | = Sheet1!$A$3 & "_" & Sheet1!$C$3 & "_" |
| 3  | = Sheet1!$A$4 & "_" & Sheet1!$C$4 & "_" |
| …  |   |
| 23 | = Sheet1!$A$24 & "_" & Sheet1!$C$24 & "_" |
| 24 | = Sheet1!$A$25 & "_" & Sheet1!$C$25 & "_" |

The corresponding values were:

|    | A |
|----|---|
| 1  | Joe_Beer_ |
| 2  | Beth_Food_ |
| 3  | Janet_Beer_ |
| …  |   |
| 23 | Janet_Car_ |
| 24 | Janet_Food_ |

I represented this "value combiner" range as another Excelsior object. The cross-tabulation referred to it rather than the original data in Harald Staff's spreadsheet. This was sufficient to construct the cross-tabulation, shown here in formula view:

| A | B | C | D |
|---|---|---|---|
| 1 | Beth | Janet | Joe |
| 2 Beer | = COUNTIF( Combine!$A$1:$A$24, "Beth_Beer_" ) | = COUNTIF( Combine!$A$1:$A$24, "Janet_Beer_" ) | = COUNTIF( Combine!$A$1:$A$24, "Joe_Beer_" ) |
| 3 Car | = COUNTIF( Combine!$A$1:$A$24, "Beth_Car_" ) | = COUNTIF( Combine!$A$1:$A$24, "Janet_Car_" ) | = COUNTIF( Combine!$A$1:$A$24, "Joe_Car_" ) |
| 4 Food | = COUNTIF( Combine!$A$1:$A$24, "Beth_Food_" ) | = COUNTIF( Combine!$A$1:$A$24, "Janet_Food_" ) | = COUNTIF( Combine!$A$1:$A$24, "Joe_Food_" ) |

**4.4 Evaluation**

To recap, I wanted to test the hypothesis that: it is possible to automate generating a table from a declarative specification of its structure, then inserting it into an existing spreadsheet to cross-tabulate its data. This was indeed possible, since the tabulation show above calculated the expected results and contained correct formulae.

Having said that, in day-to-day spreadsheet development, one would probably not need to programmatically insert just one pre-specified tabulation. I would want to find a notation that does allow users to specify table structure for themselves. This should be easy to use, but should not force the user to make the specification every time the spreadsheet is to be used, as pivot tables (Section 4.1) appear to. The stub-and-banner table examples in [StatSoft] illustrate the range of layouts this notation should describe.

Using add-ins might be better than generating "value combiner" ranges or an equivalent, especially when we need to add values to be tabulated instead of just counting them.

An obvious defect was that the spreadsheet containing the cross-tabulation, $T_s ∪ S$ in the algorithm, did not contain the cell styles, colours and so on present in the original spreadsheet $S$. This is because to form $T_s ∪ S$, $S$ was read from a saved XML file into an Excelsior object, losing the styles and other presentational information. However, as stated in Section 4.1, I was not trying to preserve these. We are now looking into the best way of doing so.

The experiment also tested the hypothesis that: not merely is it possible to automate the task of generating and inserting tabulations, but that doing so is easy using Excelsior's data structures and operations.

This was indeed easy. Firstly, independence of content from layout simplified coding, particularly when creating the table $T$.

Secondly, Excelsior represents spreadsheets as first-class values in the same way as numbers, lists, vectors, and other common structures. This permitted a functional style of programming: the algorithm could be coded as a composition of functions that returned spreadsheets and mappings. It was simple to show the code correct and to test each function separately. In contrast, most spreadsheet-manipulation systems operate directly on Excel workbooks, leading to mutable data structures and an imperative programming style, with the resulting diasdvantages familiar to computer science.

## 5. CONCLUSION

This paper has described two experiments. In one, we wrote a Prolog program that added a cross-tabulation sheet and the necessary intermediate workings to a spreadsheet. This program represented the original spreadsheet and the parts to be added using the data structures designed for representing Excelsior objects and transformations upon them. The experiment had two aims. One was to develop techniques for generating such tabulations, as an alternative to Excel pivot tables. This succeeded. The other was to assess how well suited Excelsior's representation was for this task, considered as a typical example of programmed spreadsheet manipulation. We conclude that for this task at least, they were very suitable, for the reasons given at the end of Section 4.

In the other experiment, we converted a large financial spreadsheet to Excelsior notation, then generated variants which did the same calculations, but had widely different table sizes and layout to the original. To do this conveniently, we had to invent a new notation for specifying layout to Excelsior in a way independent of table sizes. Once we had done the conversion, it took only a few minutes to generate each variant, much faster and easier than editing the original in Excel. The conversion stage was not entirely automatic, so did require care and skill. Even taking that into account, for this particular application, the net saving in time and effort was well worth it.

That experiment was a feasibility test for a business that has a large and complicated modelling spreadsheet and many clients, each needing a slightly different variant of the model. Creating all these variants in Excel and keeping them in synch with improvements to the original can be done; but it is slow and risky. To keep the original in Excelsior instead, and to derive each variant simply by changing a few size parameters or layouts, seems very promising. Together with automated testing of the resulting Excelsior modules, and ways of helping Excelsior-naïve users build client spreadsheets from these, we are exploring this work further. We are very very pleased with the results presented in this paper.